\expandafter\ifx\csname LaTeX\endcsname\relax
      \let\maybe\relax
\else \immediate\write0{}
      \message{You need to run TeX for this, not LaTeX}
      \immediate\write0{}
      \makeatletter\let\maybe\@@end
\fi
\maybe

\magnification=\magstephalf

\hsize=5.25truein
\vsize=8.3truein
\hoffset=0.37truein

\newdimen\frontindent \frontindent=.45truein
\newdimen\theparindent \theparindent=20pt


\let\em=\it

\font\tencsc=cmcsc10
\font\twelvebf=cmbx10 scaled 1200
\font\bmit=cmmib10  \font\twelvebmit=cmmib10 scaled 1200
\font\sixrm=cmr6 \font\sixi=cmmi6 \font\sixit=cmti8 at 6pt

\font\eightrm=cmr8  \let\smallrm=\eightrm
\font\eighti=cmmi8  \let\smalli=\eighti
\skewchar\eighti='177
\font\eightsy=cmsy8
\skewchar\eightsy='60
\font\eightit=cmti8
\font\eightsl=cmsl8
\font\eightbf=cmbx8
\font\eighttt=cmtt8
\def\eightpoint{\textfont0=\eightrm \scriptfont0=\fiverm 
                \def\rm{\fam0\eightrm}\relax
                \textfont1=\eighti \scriptfont1=\fivei 
                \def\mit{\fam1}\def\oldstyle{\fam1\eighti}\relax
                \textfont2=\eightsy \scriptfont2=\fivesy 
                \def\cal{\fam2}\relax
                \textfont3=\tenex \scriptfont3=\tenex 
                \def\it{\fam\itfam\eightit}\let\em=\it
                \textfont\itfam=\eightit
                \def\sl{\fam\slfam\eightsl}\relax
                \textfont\slfam=\eightsl
                \def\bf{\fam\bffam\eightbf}\relax
                \textfont\bffam=\eightbf \scriptfont\bffam=\fivebf
                \def\tt{\fam\ttfam\eighttt}\relax
                \textfont\ttfam=\eighttt
                \setbox\strutbox=\hbox{\vrule
                     height7pt depth2pt width0pt}\baselineskip=9pt
                \let\smallrm=\sixrm \let\smalli=\sixi
                \rm}


\catcode`@=11
 
\def\vfootnote#1{\insert\footins\bgroup\eightpoint
     \interlinepenalty=\interfootnotelinepenalty
     \splittopskip=\ht\strutbox \splitmaxdepth=\dp\strutbox
     \floatingpenalty=20000
     \leftskip=0pt \rightskip=0pt \parskip=1pt \spaceskip=0pt \xspaceskip=0pt
     \everydisplay={}
     \smallskip\textindent{#1}\footstrut\futurelet\next\fo@t}
 
\newcount\notenum

\def\note{\global\advance\notenum by 1
    \edef\n@tenum{$^{\the\notenum}$}\let\@sf=\empty
    \ifhmode\edef\@sf{\spacefactor=\the\spacefactor}\/\fi
    \n@tenum\@sf\vfootnote{\n@tenum}}


\tabskip1em

\newtoks\pream \pream={#\strut}
\newtoks\lpream \lpream={&#\hfil}
\newtoks\rpream \rpream={&\hfil#}
\newtoks\cpream \cpream={&\hfil#\hfil}
\newtoks\mpream \mpream={&&\hfil#\hfil}

\newcount\ncol \def\ncolp{\advance\ncol by 1}
\def\atalias#1{
    \ifx#1l\edef\xpream{\pream={\the\pream\the\lpream}}\xpream\ncolp\fi
    \ifx#1r\edef\xpream{\pream={\the\pream\the\rpream}}\xpream\ncolp\fi
    \ifx#1c\edef\xpream{\pream={\the\pream\the\cpream}}\xpream\ncolp\fi}
\catcode`@=\active

\def\taborl#1{\omit\unskip#1\hfil}
\def\taborc#1{\omit\hfil#1\hfil}
\def\taborr#1{\omit\hfil#1}
\def\multicol#1{\multispan#1\let\omit\relax}

\def\table#1\par{\midinsert\offinterlineskip\everydisplay{}
    \let@\atalias \let\l\taborl \let\r\taborr \let\c\taborc
    \def\space{\noalign{\vskip2pt}}
    \def\tablerule{\omit&\multispan{\the\ncol}\hrulefill\cr}
    \def\onerule{\space\space\tablerule\space\space}
    \def\tworules{\space\space\tablerule\space\tablerule\space\space}
    \def\annot##1\\{&\multispan{\the\ncol}##1\hfil\cr}
    \def\\{\let\\=\cr
           \edef\xpream{\pream={\the\pream\the\mpream}}\xpream
           \edef\starthalign{$$\vbox\bgroup\halign\bgroup\the\pream\cr}
           \starthalign
           \annot\hfil\tencsc Table #1\\ \noalign{\medskip}}
    \let\par\endtable}

\edef\endtable{\noalign{\vskip-\bigskipamount}\egroup\egroup$$\endinsert}

\let\plainmidinsert=\midinsert
\def\eightpttable{\def\midinsert{\let\midinsert=\plainmidinsert
    \plainmidinsert\eightpoint\tabskip 1em}\table}



\newif\iftitlepage

\def\raggedright{\rightskip 0pt plus .2\hsize\relax}

\let\caret=^ \catcode`\^=13 \def^#1{\ifmmode\caret{#1}\else$\caret{#1}$\fi}

\def\title#1\par{\vfill\supereject\begingroup
                 \global\titlepagetrue
                 \leftskip=\frontindent\parindent=0pt\parskip=0pt
                 \frenchspacing \eqnum=0
                 \gdef\runningtitle{#1}
                 \null\vskip-22.5pt\copy\volbox\vskip18pt
                 {\titlestyle#1\bigskip}}
\def\titlestyle{\raggedright\bf\twelvebf\textfont1=\twelvebmit
                \let\smallrm=\tenbf \let\smalli=\bmit
                \baselineskip=1.2\baselineskip}
\def\shorttitle#1\par{\gdef\runningtitle{#1}}
\def\author#1\par{{\raggedright#1\medskip}}

\def\shortauthor#1\par{\gdef\runningauthors{#1}}

\def\affil#1\par{{\raggedright\it#1\smallskip}}
\def@#1{\ifhmode\qquad\fi\leavevmode\llap{^{#1}}\ignorespaces}
\def\abstract{\smallskip\medskip{\bf Abstract: }}

\def\maybebreak#1{\vskip0pt plus #1\hsize \penalty-500
                  \vskip0pt plus -#1\hsize}

\def\maintextmode{\leftskip=0pt\parindent=\theparindent
                  \parskip=\smallskipamount\nonfrenchspacing}

\def\maintext#1\par{\bigskip\medskip\maintextmode\noindent}

\newcount\secnum
\def\section#1\par{\ifnum\secnum=0\medskip\maintextmode\fi
    \advance\secnum by 1 \bigskip\maybebreak{.1}
    \subsecnum=0
    \hang\noindent\hbox to \parindent{\bf\the\secnum.\hfil}{\bf#1}
    \smallskip\noindent}

\newcount\subsecnum
\def\subsection#1\par{\ifnum\subsecnum>0\medskip\maybebreak{.1}\fi
    \advance\subsecnum by 1
    \hang\noindent\hbox to \parindent
       {\it\the\secnum.\the\subsecnum\hfil}{\it#1}
    \par\noindent}

\def\references\par{\bigskip\maybebreak{.1}\parindent=0pt
    \everypar{\hangindent\theparindent\hangafter1}
    \leftline{\bf References}\smallskip}

\def\appendix#1\par{\bigskip\maybebreak{.1}\maintextmode
    \advance\secnum by 1 \bigskip\maybebreak{.1}
    \leftline{\bf Appendix: #1}\smallskip\noindent}

\def\acknowl{\medskip\noindent}

\def\bye{\endgroup\vfill\supereject\end}


\newbox\volbox
\setbox\volbox=\vbox{\hsize=.5\hsize \raggedright
       \sixit\baselineskip=7.2pt \noindent
       The Nature of Elliptical Galaxies,
       Proceedings of the Second Stromlo Symposium,
       Eds.\ M.~Arnaboldi, G.S.~Da~Costa \& P.~Saha}


\input epsf

\def\figureps[#1,#2]#3.{\midinsert\parindent=0pt\eightpoint
    \vbox{\epsfxsize=#2\centerline{\epsfbox{#1}}}
    \def\par{\endgraf\endinsert}{\bf Figure#3.}}

\def\figuretwops[#1,#2,#3]#4.{\midinsert\parindent=0pt\eightpoint
    \vbox{\centerline{\epsfxsize=#3\epsfbox{#1}\hfil
                      \epsfxsize=#3\epsfbox{#2}}}
     \def\par{\endgraf\endinsert}{\bf Figure#4.}}

\def\figurespace[#1]#2.{\midinsert\parindent=0pt\eightpoint
    \vbox to #1 {\vfil\centerline{\tenit Stick Figure#2 here!}\vfil}
    \def\par{\endgraf\endinsert}{\bf Figure#2.}}


\headline={\iftitlepage\hfil\else
              \ifodd\pageno\hfil\tensl\runningtitle
                    \kern1pc\tenbf\folio
               \else\tenbf\folio\kern1pc
                    \tensl\runningauthors\hfil\fi
           \fi}
\footline{\iftitlepage\tenbf\hfil\folio\hfil\else\hfil\fi}
\output={\plainoutput\global\titlepagefalse}


\newcount\eqnum
\everydisplay{\puteqnum}  
\def\puteqnum#1$${#1\global\advance\eqnum by 1\eqno(\the\eqnum)$$}
\def\namethiseqn#1{\xdef#1{\the\eqnum}}

 
\newcount\mpageno
\mpageno=\pageno  \advance\mpageno by 1000
 
\def\advancepageno{\global\advance\pageno by 1
                   \global\advance\mpageno by 1 }

\openout15=inx
\def\index#1{\write15{{#1}{\the\mpageno}}\ignorespaces}


\def\LaTeX{{\rm L\kern-.36em\raise.3ex\hbox{\tencsc a}\kern-.15em
    T\kern-.1667em\lower.7ex\hbox{E}\kern-.125emX}}

\def\[#1]{\raise.2ex\hbox{[}#1\raise.2ex\hbox{]}}

\def\witchbox#1#2#3{\hbox{$\mathchar"#1#2#3$}}
\def\leqsim{\mathrel{\rlap{\lower3pt\witchbox218}\raise2pt\witchbox13C}}
\def\geqsim{\mathrel{\rlap{\lower3pt\witchbox218}\raise2pt\witchbox13E}}

\def\<#1>{\langle#1\rangle}


{\obeyspaces\gdef {\ }}

\catcode`@=12 \let\@=@ \catcode`@=13
\def\+{\catcode`\\=12\catcode`\$=12\catcode`\&=12\catcode`\#=12%
       \catcode`\^=12\catcode`\_=12\catcode`\~=12\catcode`\%=12%
       \catcode`\@=0\tt}
\def\({\endgraf\bgroup\let\par=\endgraf\parskip=0pt\vskip3pt
       \eightpoint \def\/{{\eightpoint$\langle$Blank line$\rangle$}}
       \catcode`\{=12\catcode`\}=12\+\obeylines\obeyspaces}
\def\){\vskip1pt\egroup\vskip-\parskip\noindent\ignorespaces}



\title X-ray Emission from Elliptical Galaxies

\shorttitle X-ray Emission

\author Craig L. Sarazin

\shortauthor Sarazin

\affil Department of Astronomy, University of Virginia, U.S.A.

\abstract

Elliptical galaxies are generally luminous sources of X-ray radiation,
and contain large amounts of hot, interstellar gas.
In the brighter X-ray galaxies, the inferred masses of hot gas are
consistent with those expected given the present rates of stellar mass loss.
The required rates of heating of the gas are also roughly consistent with
those expected from the motions of gas losing stars.
X-ray observations, particularly X-ray spectra, require a low rate of
Type Ia supernova heating and chemical enrichment in the gas.
\index{supernovae}
In the brightest X-ray galaxies, the cooling times in the gas are short,
which suggests that the gas forms steady-state cooling flows.
\index{cooling flow}
Steady cooling models explain most of the properties of the brighter X-ray
galaxies, including their luminosities, the X-ray--optical correlation,
their temperatures, and their surface brightness profiles.
Although the optical and X-ray luminosities of early-type galaxies are
strongly correlated, there is a large dispersion in this correlation.
The origin of the emission in the X-ray faint ellipticals is less
certain.
All ellipticals appear to have a hard X-ray spectral component due to
accreting binary systems.
X-ray faint ellipticals also have a very soft X-ray component, which
may be residual hot interstellar gas.
The X-ray spectra of ellipticals indicate that the abundance of iron
is well below the solar value, which implies that the rate of Type Ia
supernova contamination is small.
The abundances and evidence for gradient gradients suggest that
stellar abundance gradients and inflow of the gas affect the X-ray
spectra.

\section Introduction

\index{X-ray emission}
One of the most important discoveries of the
$Einstein$ X-ray Observatory was that normal elliptical are
generally strong X-ray sources
(Forman et al.\ 1985;
Trinchieri \& Fabbiano 1985).
The X-ray emission indicates that these galaxies contain
extensive atmospheres of hot, diffuse interstellar gas.
Prior to this discovery, it was generally believed that
early-type galaxies were gas-poor systems
(Faber \& Gallagher 1976).
Surveys of elliptical galaxies based on $Einstein$ data include
Nulsen et al.\ (1984),
Forman et al.\ (1985),
Canizares et al.\ (1987),
and
Fabbiano et al.\ (1992).
A number of $Ginga$ observations of early-type galaxies
are analyzed in Awaki et al.\ (1991).
A recent compilation of ROSAT observations is given in
Davis \& White (1996).
ASCA results on the X-ray spectra of ellipticals described in
Awaki et al.\ (1994),
Loewenstein et al.\ (1994),
Matsushita et al.\ (1994),
Mushotzky et al.\ (1994),
and
Matsumoto et al.\ (1996).

\section X-ray Properties

Elliptical galaxies have X-ray luminosities which range over
$L_X \approx 10^{39} - 10^{42}$ erg s$^{-1}$.
(All comparisons to observations in this paper
assume a Hubble constant $H_o = 50$ km/s/Mpc, and a distance to
the
center of the Virgo cluster of 25 Mpc.)
There appears to be a strong correlation between the X-ray and blue optical
luminosities of early-type galaxies, with $L_X \propto L_B^{1.6-2.3}$.
Figure 1 shows the X-ray and optical luminosities of the
early-type galaxies in the survey of Canizares et al.\ (1987).
The filled circles are detections, while the inverted triangles are upper
limits on $L_X$.
Obviously, there is considerable scatter about this relation.

The X-ray emission from ellipticals is extended, with typical
overall sizes of $R_X \sim 50$ kpc.
For example, Figure 2 shows the ROSAT X-ray image of NGC~4472 from
Irwin \& Sarazin (1996).
\index{NGC 4472}
On large scales, the azimuthally-averaged X-ray surface
brightnesses $I_X (r)$ of ellipticals typically decline with
radius $r$ roughly as $I_X \propto r^{-2}$
(Forman et al.\ 1985).
Very crudely, the X-ray and optical surface brightnesses of
ellipticals decline in proportion to one another
$I_X (r) \propto I_B (r)$
(Trinchieri et al.\ 1986).
This relationship holds only very approximately;
the X-ray surface brightness profiles of ellipticals often have
structure which is not apparent in the optical profiles.
Also, this proportionality holds only within a given elliptical.
The constant of proportionality varies from galaxy to galaxy,
as required by the steeper than linear correlation between
X-ray and optical luminosities.

\figureps[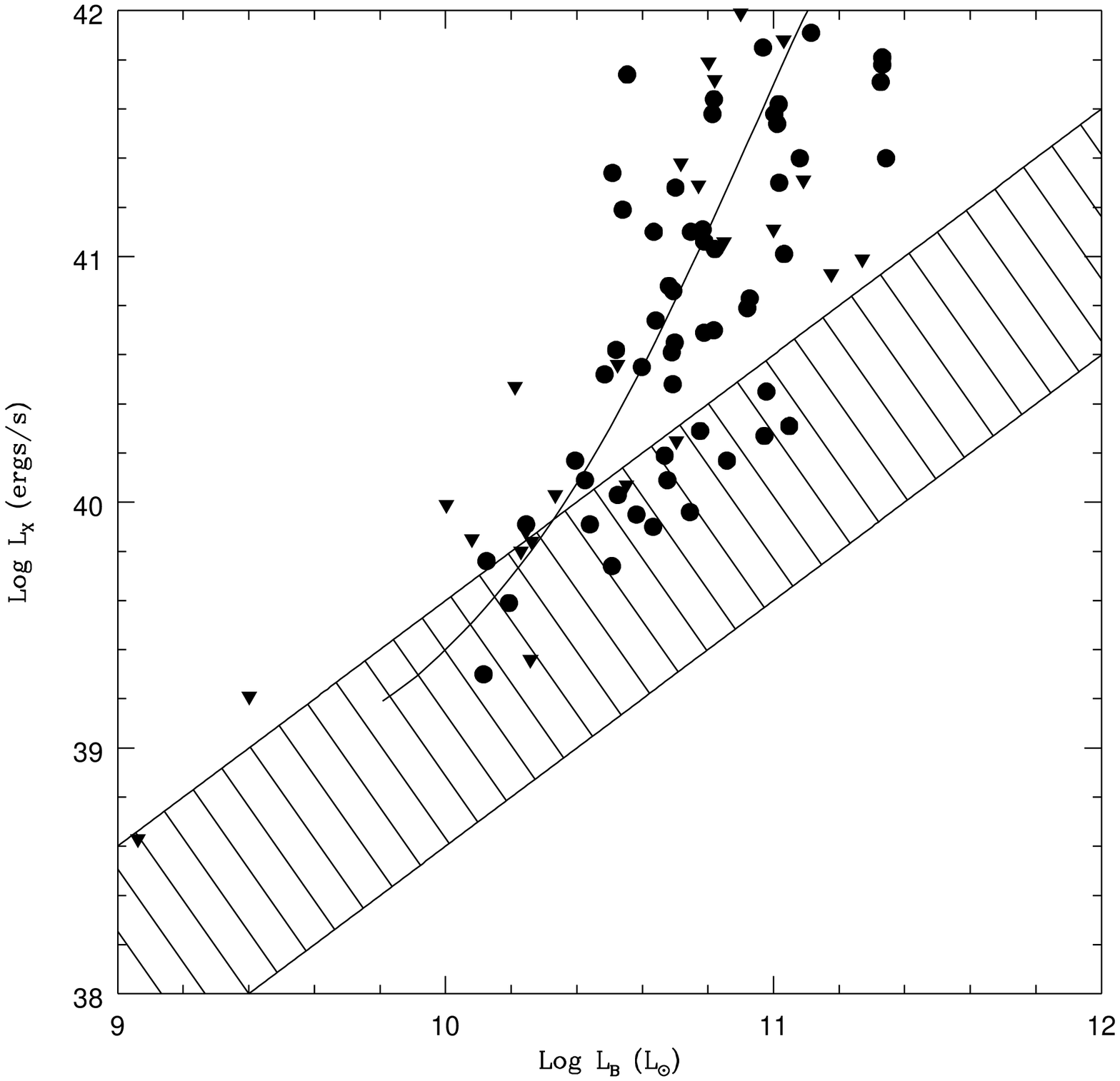,.6\hsize] 1.  The correlation
of X-ray and optical luminosities of early-type
galaxies from Canizares et al.\ (1987).
The filled circles are detections, and the inverted triangles are
upper limits.
The hatched area gives the range of estimated stellar X-ray
luminosities from
Forman et al.\ (1985),
Canizares et al.\ (1987),
and
Fabbiano et al.\ (1989).
The curve gives the predicted relation for the steady cooling models
of Sarazin \& Ashe (1989).
\index{binary stars}
\index{cooling flow}

At least for the more luminous X-ray ellipticals,
spectral observations indicate that the X-rays are produced
by thermal emission from diffuse gas.
The spectra imply that the temperature of the gas is
typically $T \approx 10^7$ K ($kT \approx 1$ keV)
(Forman et al.\ 1985;
Trinchieri et al.\ 1986).
The X-ray surface brightnesses profiles of ellipticals
then indicate that the hot gas density declines with
radius at large radii roughly as
$\rho_{gas} \propto r^{-3/2}$.
The total gas masses required by the X-ray observations
are rather uncertain, because this density distribution
doesn't converge and the X-ray images of ellipticals fade
into the background at large radii.
However, the gas masses are often in the range
$M_{gas} \approx 10^9 - 10^{11} \, M_\odot$.
It is useful to compare the gas masses to the optical
luminosities of the galaxies.
For the brighter ellipticals, one finds
$( M_{gas}/M_\odot ) \approx 0.2 ( L_B / L_\odot )$.
If a stellar mass to light ratio of 
$\approx$$10 \, M_\odot / L_\odot$
is assumed, then is implies that the gas mass is about
2\% of the stellar mass.
The hot gas is generally the dominant form of interstellar matter
in elliptical galaxies (by mass).
The interstellar gas masses in brighter ellipticals are large, but
represent a much smaller fraction of the stellar mass
than in late-type spiral galaxies.

\figureps[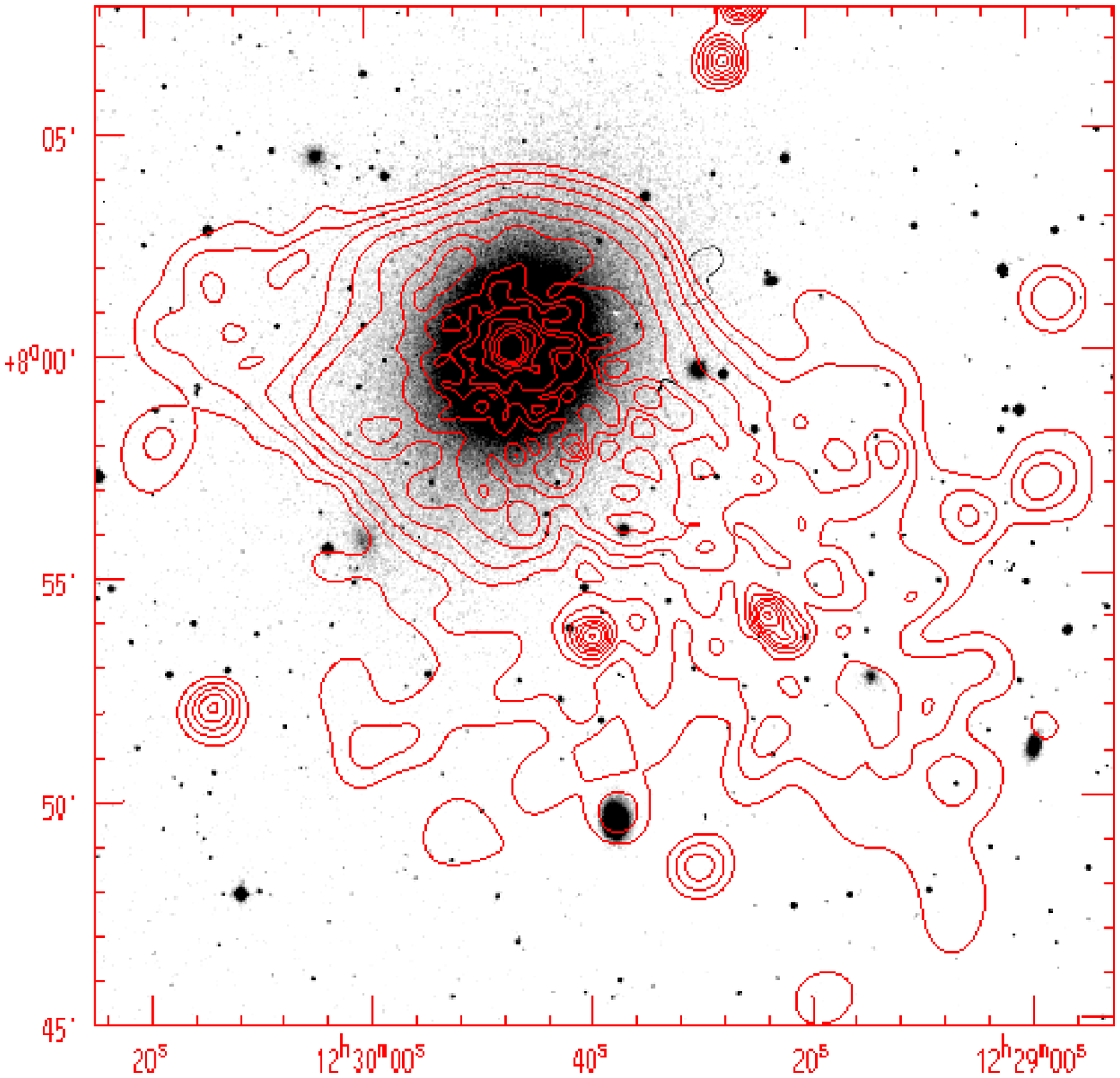,.6\hsize] 2.  The ROSAT
X-ray image of the Virgo elliptical NGC~4472
(Irwin \& Sarazin 1996).
Contours of the X-ray emission are superposed on
a greyscale representation of the optical image.

\section Physical State of the Gas

\subsection Origin of the Gas

The amount of hot gas seen in the brighter X-ray ellipticals is
consistent with the amount expected for the present rates of
stellar mass loss acting over a significant fraction of the
Hubble time.
The present rate of stellar mass loss in early-type galaxies is
fairly well-determined by stellar evolution theory to be
$ ( {\dot M}_* / L_B ) \approx 1.5 \times 10^{-11} \, M_\odot
\, {\rm yr}^{-1} \, L_\odot^{-1}$
(Renzini \& Buzzoni 1986).
If all of the stellar mass loss is converted into hot gas and
the same rates have applied for a time $t$,
the ratio of the mass of hot gas to the optical luminosity would now be
$( M_{gas}/M_\odot ) \approx 0.16 ( L_B / L_\odot )
(t / 10^{10} \, {\rm yr} )$, which is quite close to the
typical observed value for bright ellipticals.

There are several caveats that need to be given concerning this agreement
of the observed and predicted gas masses.
First, models for the stellar evolution of elliptical galaxies predict that
the stellar mass loss rate was much higher in the past.
This and the fact that the gas in clusters of galaxies contains a very
large mass of heavy elements suggest that elliptical galaxies have lost
substantial amounts of gas in the past.
Second, the agreement of the present stellar mass lost rate and the
amount of hot gas really only applies to the brightest X-ray galaxies.
For the fainter X-ray galaxies in Figure 1, the masses of hot gas are
apparently much smaller than from the present rates of stellar mass
loss.
Third, it is not obvious that all the gas resulting from stellar
mass loss will be heated to X-ray emitting temperatures.
Finally, elliptical galaxies are seldom very isolated, and
large ellipticals near
the centers of groups or poor clusters may be accreting ambient gas
(Thomas 1986).

\subsection Heating of the Gas

Why is the interstellar gas in ellipticals hot?
The primary reason is that the stars in elliptical galaxies
move at large velocities relative to nearby stars.
Thus, gas which is ejected from one star (for example,
in a red giant wind or planetary nebula) will encounter gas
ejected from other stars at a very high velocity,
determined by the stellar velocity dispersion of the
galaxy.
Collisions between the ejected gas and either gas ejected by
neighboring stars or ambient hot gas may thermalize the kinetic
energy of motion of the gas.
Because this gives an amount of heat which is proportional
to the amount of gas injected, 
it is useful to define the energy per unit injected gas mass as
$3 k T_{inj} / 2 \mu m_p$.
The resulting heating due to stellar mass loss is then
$$
T_{inj} = {{\mu m_p \sigma_*^2}\over{k}}
= 6.8 \times 10^6 \, {\rm K} \,
\left( {{\sigma_*}\over{300 \, {\rm km \, s}^{-1}}} \right)^2 ,
$$
\namethiseqn\eqtinjstar
where $\sigma_*$ is the one-dimensional stellar velocity dispersion,
and $\mu$ is the mean mass per particle in the gas in units of
the proton mass $m_p$.
This gives a temperature which is comparable to those observed
in ellipticals.
If the gas in ellipticals forms a cooling flow, there will be
additional heating due to infall in the galactic potential and
adiabatic compression.
\index{cooling flow}
Since the stellar velocity dispersion measures the depth of the potential
well in an elliptical galaxy, this source of heating is also proportional
to $\sigma_*^2$, and can be viewed as a proportionate increase in
equation \eqtinjstar.

In addition to the relatively quiescent stellar mass loss
which dominates the injection of mass into the interstellar
gas, a small fraction of the gas may be ejected with large
kinetic energy by Type Ia supernovae.
\index{supernovae}
The heating rate per unit injected gas mass due to supernovae is
$$
T_{inj} = 3.6 \times 10^7 \, {\rm K} \,
\left( {{r_{SN}}\over{0.22}} \right)
\left( {{E_{SN}}\over{10^{51} \, {\rm ergs}}} \right)
\left( {{{\dot M}_* / L_B}\over{1.5 \times 10^{-11}} \, M_\odot
\, {\rm yr}^{-1} \, L_\odot^{-1}} \right)^{-1} \, .
$$
\namethiseqn\eqtinjsn
Here,
$E_{SN}$ is the average kinetic energy of each supernova ejection,
and the rate of supernovae is $r_{SN}$ per $10^{10} \, L_\odot$ (in blue
luminosity) per century.
Tammann (1982) gives $r_{SN} = 0.22$, but more recent surveys give
lower values
(e.g., Evans et al.\ 1989).

Comparing equations \eqtinjstar\ and \eqtinjsn, it is clear that
supernova heating would dominate in all ellipticals if the supernova
rate is as high as that given by Tammann (1982).
\index{supernovae}
However, there are many reasons why the X-ray observations of ellipticals
are inconsistent with such a high supernovae heating rate.
First, this much heating would cause many ellipticals to have
winds, which are not observed.
Second, the predicted X-ray luminosities would be too large in most
cases if this much supernova heating were converted into X-ray
emission.
Finally, the heavy element abundances in the hot gas in ellipticals would
be very high if the supernova rate was high, whereas they are observed
to be quite small (\S~7).

\subsection Cooling of the Gas

One certain mechanism for energy loss by the hot gas is the emission of
the observed X-rays.
X-ray emission by a hot, thin plasma is the result of processes involving
electron-ion collisions (e.g., thermal bremsstrahlung and line emission),
and the emissivity is proportional to the square of the density.
One can write the emissivity or the cooling rate per unit volume of
the gas as $\rho^2 \Lambda ( T )$, where $\rho$ and $T$ are the
gas density and temperature, respectively.
In order to assess whether this cooling is likely to affect the gas
significantly, it is useful to estimate the time required to cool
the gas completed if these energy losses are not balanced.
The total isobaric cooling time of the gas is given by
$$
t_{cool} = {5\over2} \, {{1}\over{P}}
\int_0^\theta {{\theta \, d \theta}\over{\Lambda ( \theta )}} \, ,
$$
\namethiseqn\eqtcool
where $\theta \equiv k T / ( \mu m_p )$.
In general, the cooling time increases with radius in an elliptical
galaxy.
However, for the brighter X-ray ellipticals, the cooling time is
shorter than $10^{10}$ yr over essentially all of the galaxy
(see Table 1 in Sarazin [1990]).
The major sources of heating of the gas are associated with its
injection
(equations \eqtinjstar\ and \eqtinjsn\ above),
and cannot balance continuous cooling.
\index{cooling flow}
Thus, one expects the gas to cool, and approach steady-state
cooling over most of the observed regions of the brighter galaxies.

\section Evolution of the Gas in Ellipticals

There are several reasons why the interstellar gas in ellipticals
should evolve over time.
First, the time scales for cooling and inflow become 
longer than the age at large radii, and the gas flows
there are presumably time dependent.
Second, the rates of stellar mass loss and supernova heating vary with time.
\index{supernovae}

Spherically symmetric, time-dependent hydrodynamical
simulations of elliptical galaxies have been made by a number
of authors, including
Loewenstein \& Mathews (1987),
D'Ercole et al.\ (1989),
and
David et al.\ (1990).
Most of these models assume that elliptical galaxies form in a single
short burst of star formation.
Many of the models follow the evolution only after this burst.
The most important characteristic of these models tends to be
the rate of variation of the supernova heating rate compared to
the total rate of stellar mass loss.
In general, most models have a supernova rate which was higher
in the past;
this is particularly true if one includes the initial burst of
star formation and Type II supernovae.
\index{cooling flow}

In the models where the supernova heating rate was very high
in the past, the gas initially forms a transonic wind.
As the rate of supernovae declines, these models may undergo
a period of subsonic inflation.
The heating of the gas is insufficient to unbind it from the galaxy, and
most of the injected energy goes into increasing the pressure in the gas
and causing it to inflate slowly.
Finally, once the cooling time becomes shorter than the age of the
galaxy, a steady-state cooling flow forms.
In this phase, the X-ray luminosity of the galaxy is essentially equal
to the total rate of heating of the gas.
In \S~3.3, we noted that the cooling time is less
than the age over essentially the entire observed regions of hot
gas in the brighter X-ray ellipticals.
Thus, these galaxies are likely to be undergoing steady-state
cooling at present.

\section Steady-State Cooling

One-dimensional, steady-state, cooling flow models for the gas in
ellipticals have been calculated by
Thomas (1986),
Sarazin \& White (1987),
Vedder et al.\ (1988),
Sarazin \& Ashe (1989),
and others.
In these models, the rate of heat input associated with the injection
of gas by stellar mass loss equal the rate of emission of the gas
as it cools.
Thus, the X-ray luminosity for steady cooling is given approximately by
$$
L_X \approx
\left( {{3 k T_{inj}}\over{2 \mu m_p}} \right)
\dot M_* \, .
$$
\namethiseqn\eqlx
Steady cooling models give X-ray luminosities which agree with those
of the brighter X-ray ellipticals, as long as the supernova heating
rate is low.
\index{cooling flow}

These models also provide a simple explanation for the steep dependence
of the X-ray luminosity on the optical luminosity.
If the supernova rate is low, then the injection temperature of the
gas varies with the square of the velocity dispersion
(equation \eqtinjstar).
According to the Fundamental Plane or Faber--Jackson relations, the
velocity dispersion of an elliptical increases with its luminosity.
When combined with equations \eqtinjstar\ and \eqlx, this leads
to $L_X \propto L_B^{1.5-2}$, consistent with the observed relation.
In Figure 1, the predicted $L_X$--$L_B$ relationship is shown
for detailed cooling flow models from Sarazin \& Ashe (1989).

Steady cooling models give a reasonably good fit to the radial X-ray
surface brightness profiles of ellipticals.
There is a simple argument which explains this (Sarazin 1986).
In steady cooling models, the rate of heating of the gas equals
the rate of cooling through the emission of X-rays
(equation \eqlx).
If this applies locally as well as globally, then the emissivity of
the gas equals the local heating rate per unit volume.
If one integrates the X-ray emissivity along a line of sight
through the galaxy, this gives
$$
I_X (r) \approx {{3}\over{2}} \, {{k}\over{\mu m_p}} \,
\langle  T_{inj} \rangle 
\left( {{{\dot M}_*}\over{L_B}} \right)
\left< {{\Lambda_X}\over{\Lambda}} \right> \, I_B (r) \, ,
$$
\namethiseqn\eqix
where the averages are along the line-of-sight at $r$.
The factor $\langle \Lambda_X / \Lambda \rangle$ is a bolometric correction
for the fraction of the gas emission is the X-ray band, which is nearly
unity.
If the velocity dispersion of the galaxy is relatively constant,
the $\langle  T_{inj} \rangle$, then equations \eqtinjstar\ and
\eqix\ imply that $I_X \propto I_B$, approximately as is observed.
Also, if one adopts a Hubble law for the optical surface brightness of
ellipticals, then this gives $I_X \propto r^{-2}$
\index{cooling flow}

\figureps[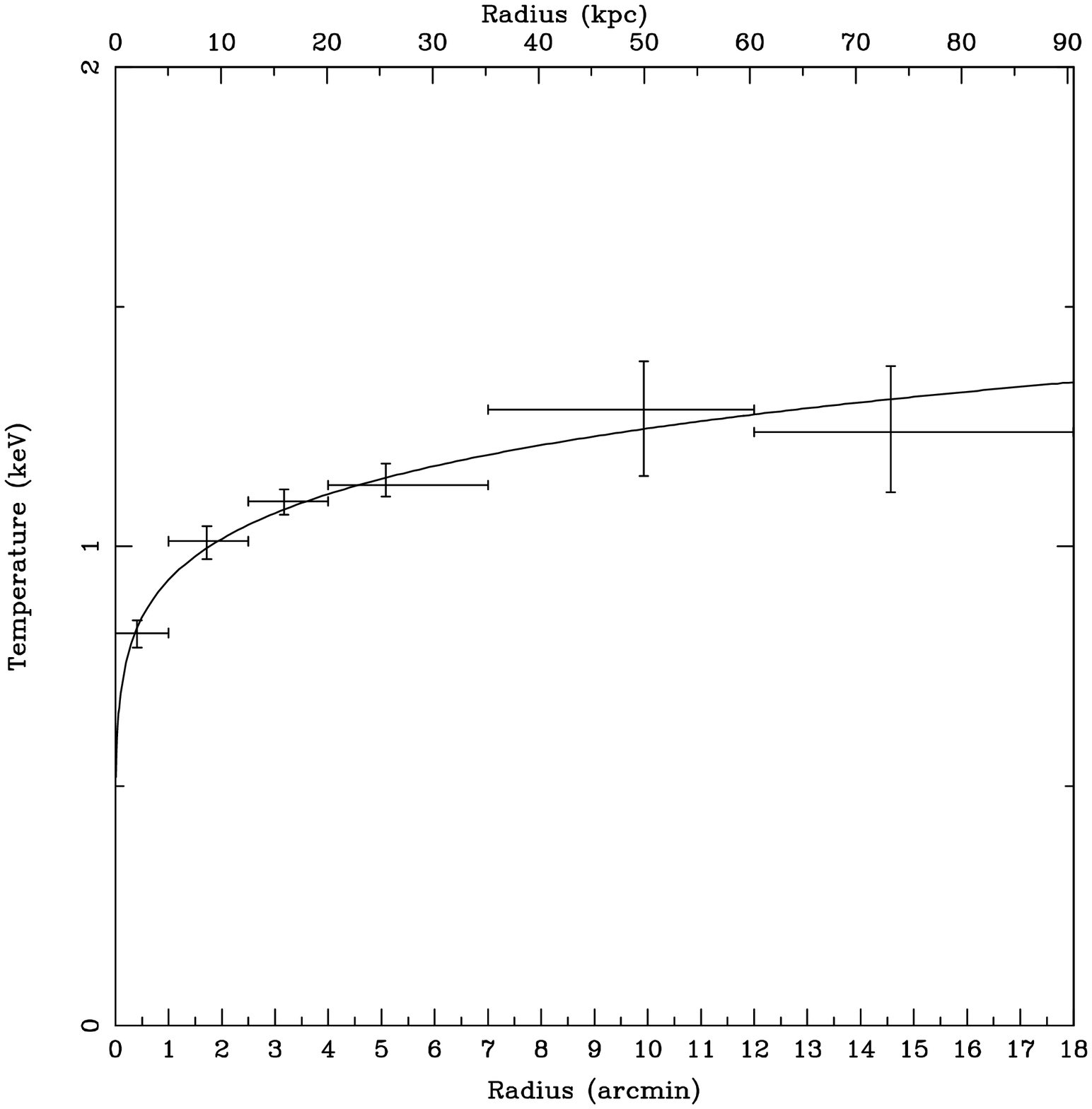,.5\hsize] 3.  The radial variation of the
temperature in NGC~4472 from the ROSAT PSPC X-ray spectrum
(Irwin \& Sarazin 1996).
The curve is a fit to the data.

Finally, the steady cooling models are consistent with the temperatures
and temperature variations observed in the brighter X-ray ellipticals.
The steady cooling models generally predict that the gas temperatures
are relatively constant in the outer regions, and decline toward the
center.
ROSAT and ASCA spectra show this pattern in all of the bright
ellipticals of which I am aware.
For example, Figure 3 shows the observed temperature profile
of NGC~4472
(Irwin \& Sarazin 1996).
\index{NGC 4472}

\section X-ray Faint Ellipticals

Most of the detailed X-ray properties of ellipticals discussed above
are based on observations of the brightest ellipticals.
However, it is clear from Figure 1 that there is a range of
at least an order of magnitude in the X-ray luminosities
of ellipticals with the same optical luminosity.
Why is there such a large dispersion in X-ray luminosities?
What is the origin of the detected X-ray emission of the fainter
ellipticals?
Is it thermal emission by interstellar gas as in the brighter ellipticals?

There have been a number of suggestions as to the origin in the dispersion
in the X-ray luminosities of ellipticals.
D'Ercole et al.\ (1989) suggested that the faint X-ray ellipticals are at
an earlier evolutionary stage and undergoing subsonic inflation
(\S~4).
However, the specific version of this hypothesis which they advocated
required a large supernova heating rate, which may be ruled out by the
X-ray spectra of ellipticals and other X-ray data
(\S\S~3.2, 7).
\index{supernovae}

White \& Sarazin (1991) suggested that the X-ray faint ellipticals were
mainly located in denser regions, and that the gas in these galaxies had
been removed by ram pressure stripping.
While there are clear cases of this occurring
(e.g., M~86 in the Virgo cluster),
\index{NGC 4406}
the statistical anticorrelation of X-ray luminosity with local density is
not very well established
(Eskridge et al.\ 1995).

What is the nature of the X-ray emission in the X-ray faint ellipticals?
It might be due to residual diffuse hot ISM, in which case it would have
a soft X-ray spectrum, or due to low mass X-ray binaries [LMXRBs],
with a hard X-ray spectrum.
In Figure~1, the shaded area shows the predicted X-ray luminosities of
due to LMXRBs based on a number of different extrapolations from nearby
galaxies.
Because the stellar populations of bright ellipticals appear to be fairly
homogeneous, one expects the $L_X$ contribution of LMXRBs to scale with the
optical luminosity $L_B$.

ASCA X-ray spectra of essentially all ellipticals show the presence of
a very hard spectral component which is approximately proportional to
the optical luminosity of the galaxy
(Matsushita et al.\ 1994;
Matsumoto et al.\ 1996).
This hard spectral component is consistent in spectral shape and flux with
that expected from LMXRBs.
So, it is likely that binary stars are an important source of X-ray
emission in X-ray faint ellipticals.
\index{binary stars}

However, ROSAT and ASCA spectra also show evidence for a very soft X-ray
spectral component in X-ray faint ellipticals
(Fabbiano et al.\ 1994).
This component might be due to diffuse ISM, to stellar corona, or to
some unknown cause,
although there is some evidence favoring an interstellar origin
(Davis \& White 1996).

Thus, it seems likely that the X-ray emission of X-ray faint ellipticals
is due to several mechanisms, including LMXRBs and residual hot ISM.

\section Abundances in the X-ray Gas

At the temperatures found in elliptical galaxies, much of the X-ray
emission is line emission due to heavy elements.
\index{abundances}
Thus, X-ray spectra can be used to determine the abundances in the gas,
particularly the abundance of iron.
It was expected that the gaseous abundances would be moderately to
extremely high, depending on the supernova rate.
\index{supernovae}
The heavy element abundances in stars in the central regions of elliptical
galaxies are rather high, and Type Ia supernovae may further enhance the
abundances.
The expected average gaseous iron abundance is then
$$
{{({\rm Fe / H})_{gas}}\over{({\rm Fe / H})_\odot}}
\approx \left[
{{({\rm Fe / H})_*}\over{({\rm Fe / H})_\odot}} +
1.6
\left( {{M_{Fe}}\over{0.6 \, M_\odot}} \right)
\left( {{r_{SN}}\over{0.075}} \right) \right] \, ,
$$
\namethiseqn\eqfe
where $({\rm Fe / H})_*$ is the iron abundance in the normal
stellar ejecta, and each Type Ia supernova produces a mass
$M_{Fe}$ of iron.
The supernova rate used in equation \eqfe\ is at the low end of the
suggested rates, so the iron abundances in ellipticals were expected to be
$({\rm Fe / H})_{gas} \geqsim 3 ({\rm Fe / H})_\odot$,
and could reach ten times solar if higher
supernova rates apply.

The expected gaseous abundances in ellipticals are also sensitive to
stellar abundance gradients and the dynamical state of the gas.
The stellar abundance gradients observed in elliptical galaxies
will produce gradients in the interstellar abundances, particularly
for elements which are not provided mainly by Type Ia supernovae.
If the gas in involved in an outflow (e.g., D'Ercole et al.\ 1989),
the heavy elements at any radius will have come from stars interior to
that radius, and this will increase the abundances.
Conversely, if the gas is involved in an inflow as in a cooling flow,
all of the heavy elements come from the  outer regions of the galaxy,
and the abundances are depressed.

X-ray spectra from ROSAT and particularly ASCA show that the interstellar
abundances of iron are quite low,
$({\rm Fe / H})_{gas} \approx 0.1 - 0.5 ({\rm Fe / H})_\odot$
(Awaki et al.\ 1994;
Loewenstein et al.\ 1994;
Mushotzky et al.\ 1994;
Matsumoto et al.\ 1996).
In a number of cases, abundance gradients are seen.
Such a low iron abundance requires that the rate of Type Ia supernovae
mass injection must be very low
(equation~\eqfe).
The low abundances and the presence of abundance gradients suggests that
stellar abundances gradients and inflow affect the spectra.
However, the observed abundances are so low that external inflow
of low abundance gas or incomplete mixing of stellar ejecta into
the hot gas may be required.
It would be very helpful in this regard if accurate stellar iron abundance
gradients could be determined for many of the X-ray bright ellipticals.

\references

Awaki, H., Koyama, K., Kunieda, H., Takano, S., Tawara, Y., \& Ohashi, T.
	1991, ApJ, 366, 88

Awaki, H., et al., 1994, PASJ, 46, L65

Canizares, C. R., Fabbiano, G., \& Trinchieri, G. 1987, ApJ, 312, 503

David, L. P., Forman, W., \& Jones, C. 1990, ApJ, 359, 29

Davis, D. S., \& White, R. E., III, 1996, preprint

D'Ercole, A., Renzini, A., Ciotti, L., \& Pellegrini, S. 1989, ApJ, 341, L9

Eskridge, P., Fabbiano, G., \& Kim, D.-W. 1995, ApJS, 97, 141

Evans, R., van den Bergh, S., \& McClure, R. 1989, ApJ, 345, 752

Fabbiano, G., Gioia, I., \& Trinchieri, G. 1989, ApJ, 347, 127

Fabbiano, G., Kim, D.-W., \& Trinchieri, G. 1992, ApJS, 80, 531

Fabbiano, G., Kim, D.-W., \& Trinchieri, G. 1994, ApJ, 429, 94

Faber, S. M., \& Gallagher, J. S. 1976, ApJ, 204, 365

Forman, W., Jones, C., \& Tucker, W. 1985, ApJ, 293, 102

Irwin, J. A., \& Sarazin, C. L. 1996, ApJ, in press

Loewenstein, M., \& Mathews, W. G. 1987, ApJ, 319, 614

Loewenstein, M., Mushotzky, R., Tamura, T., Ikebe, Y., Makishima, K.,
	Matsushita, K., Awaki, H., \& Serlemitsos, P. 1994, ApJ, 436, L75

Matsumoto, H., Koyama, K., Awaki, H., Tsuru, T., Loewenstein, M., \&
	Matsushita, K. 1996, preprint

Matsushita, K., et al., 1994, ApJ, 436, L41

Mushotzky, R., Loewenstein, M., Awaki, H., Makishima, K., Matsushita, K.,
	\& Matsumoto, H. 1994, ApJ, 436, L79

Nulsen, P. E., Stewart, G. C., \& Fabian, A. C. 1984, MNRAS, 208, 185

Renzini, A., \& Buzzoni, A. 1986, in Spectral Evolution in Galaxies,
	ed.\ C. Chiosi and A. Renzini (Dordrecht: Reidel), 195

Sarazin, C. L. 1986. in Gaseous Halos around Galaxies, ed.\ J. Bregman \&
	F. Lockman (Greenbank: NRAO), 223

Sarazin, C. L. 1990, in The Interstellar Medium in Galaxies,
	ed.\ H. A. Thronson, Jr., \& J. M. Shull (Dordrecht: Kluwer), 201

Sarazin, C. L., \& Ashe, G. A. 1989, ApJ, 345, 22

Sarazin, C. L., \& White, R. E., III, 1987, ApJ, 320, 32

Tammann, G. A. 1982, in Supernovae: A Survey of Current Research,
	ed.\ by M. J. Rees \& R. J. Stoneham (Dordrecht: Reidel), 371

Thomas, P. A. 1986, MNRAS, 220, 949

Trinchieri, G., \& Fabbiano, G. 1985, ApJ, 296, 447

Trinchieri, G., Fabbiano, G., \& Canizares, C. R. 1986, ApJ, 310, 637

Vedder, P. W., Trester, J. J., \& Canizares, C. R. 1988, ApJ, 332, 725 

White, R. E., III, \& Sarazin, C. L. 1991, ApJ, 367, 476

\acknowl
The work was supported by NASA Astrophysical Theory Program grant
NAG 5-3057, NASA ASCA grant NAG 5-2526, and NASA ROSAT grant 5-3308.

\bye